\documentclass[journal]{IEEEtran}

\usepackage[cmex10]{amsmath}
\usepackage{amssymb}
\usepackage{epsfig}
\usepackage{subfigure}
\usepackage{cite}
\usepackage{graphicx}
\usepackage{color}
\usepackage{bm}
\usepackage{booktabs}
\usepackage{gensymb}
\usepackage{mathrsfs}
\usepackage{xfrac}
\usepackage{algorithm}
\usepackage{algorithmic}
\newlength{\figwidth}
\newtheorem{lemma}{\it Lemma}

\makeatletter
\renewcommand\normalsize{%
  \@setfontsize\normalsize\@xpt\@xiipt
   \abovedisplayskip 5\p@ \@plus4 \p@ \@minus5\p@
   \abovedisplayshortskip \z@ \@plus4\p@
   \belowdisplayshortskip 5\p@ \@plus4\p@ \@minus5\p@
   \belowdisplayskip \abovedisplayskip
   \let\@listi\@listI}
\makeatother

\begin{document}

\title{Physical Layer Security in Near-Field Communications}
\author{Zheng Zhang, Yuanwei Liu, Zhaolin Wang, Xidong Mu, and Jian Chen \vspace{-10mm}
\thanks{Zheng Zhang and Jian Chen are with the School of Telecommunications Engineering, Xidian University, Xi'an 710071, China (e-mail: zzhang\_688@stu.xidian.edu.cn; jianchen@mail.xidian.edu.cn). Yuanwei Liu, Zhaolin Wang and Xidong Mu are with the School of Electronic Engineering and Computer Science,
Queen Mary University of London, London E1 4NS, U.K. (e-mail: yuanwei.liu@qmul.ac.uk; zhaolin.wang@qmul.ac.uk; xidong.mu@qmul.ac.uk;).}
}
\maketitle

\begin{abstract}
    A near-field secure transmission framework is proposed. Employing the hybrid beamforming architecture, a multi-antenna base station (BS) transmits confidential information to a multi-antenna legitimate user (U) against a multi-antenna eavesdropper (E) in the near field. A two-stage algorithm is proposed to maximize the near-field secrecy capacity. Based on the fully-digital beamformers obtained in the first stage, the optimal analog beamformers and baseband digital beamformers can be alternatingly derived in the closed-form expressions in the second stage. Numerical results demonstrate that in contrast to the far-field secure communication relying on the \emph{angular disparity}, the near-field secure communication mainly relies on the \emph{distance disparity} between U and E.
\end{abstract}

\vspace{-4mm}
\IEEEpeerreviewmaketitle

\section{Introduction}
To fulfill the growing demands for the ubiquitous connectivity of the sixth generation (6G) wireless communications, tremendous efforts have been devoted to devising emerging technologies, e.g., millimeter wave (mmWave), terahertz (THz), and ultra-massive multiple-input-multiple-output (UM-MIMO) \cite{W.Saad_6G}. However, all these key enablers rely on the employment of large-scale antennas and high frequencies, which inevitably causes wireless communications to be operated in the near-field region. In contrast to the conventional \textit{planar-wave} channel model of far-field scenarios, electromagnetic (EM) propagation is accurately characterized by the \textit{spherical-wave} channel model \cite{H.Zhang_NF,C.Huang_NF2} in near-field communications. The unique spherical-wave propagation model contains both the direction and distance information of the receiver, which makes array radiation patterns focus on a specific point (i.e., \textit{beam focusing}) of the free space. Thus, near-field communications can utilize the new dimension of distance to achieve more precise signal enhancement and interference management for wireless networks, which has drawn a wide range of attention recently \cite{C.Huang_NF1,J.Xu_STAR-RIS,L.Dai_NF1}.

\begin{figure}[!t]
 \centering
 \subfigure[The far-field secure communication using beam steering.]{
  \includegraphics[scale = 0.26]{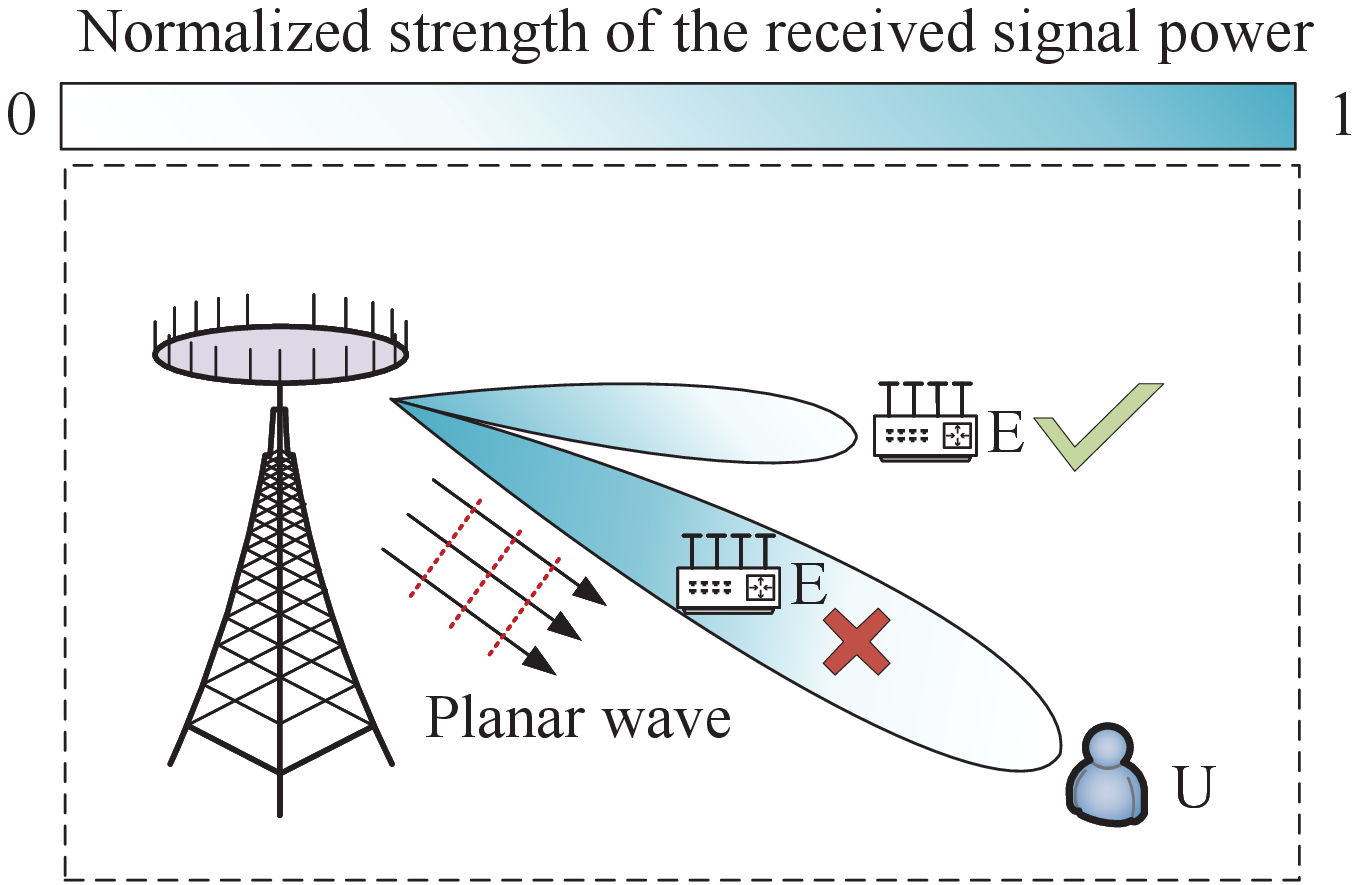}
   \label{fig1:(a)}
   }
\subfigure[The near-field secure communication using beam focusing.]{
 \includegraphics[scale = 0.26]{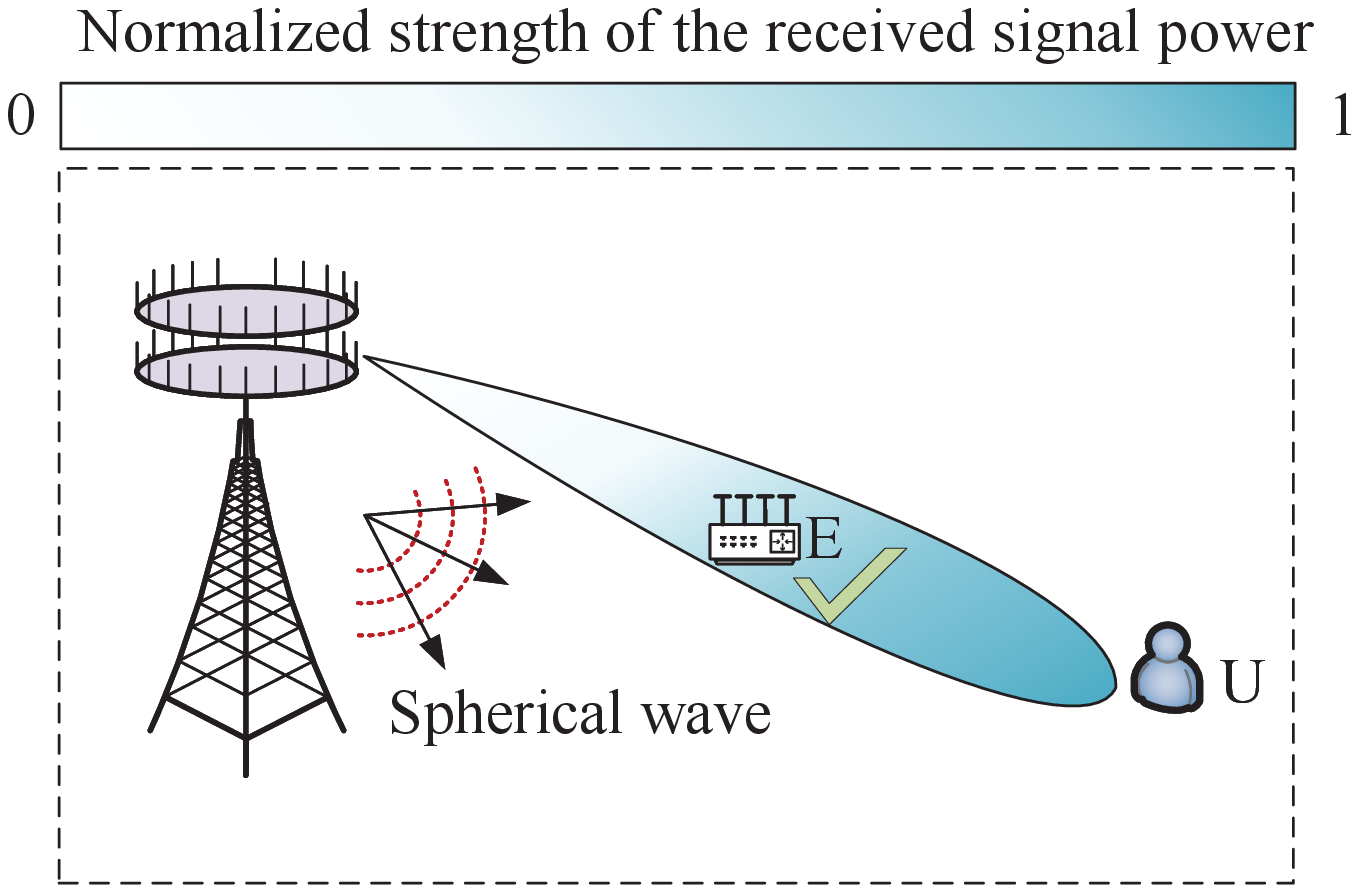}
 \label{fig1:(b)}
}
\caption{Comparison of secure transmission in far-field and near-field networks.}
\label{fig:tran1}
\end{figure}

Due to the broadcast characteristics of wireless channels, the transmitted signal is exposed to vulnerable environments and is easily wiretapped by the malicious eavesdropper (E). As a complement to cryptography, physical layer security (PLS) is proposed to safeguard private information from eavesdropping \cite{M.Bloch_PLS}. PLS is capable of exploiting the physical characteristics of wireless channels, e.g., interference, fading, noise, directivity, and disparity, without introducing complicated secret key generation and management. Nevertheless, most works for PLS mainly focused on the planar-wave channel model of the far field \cite{Y.Liu_NOMA_PLS,ZZ_STAR,Q.Shi_MIMO_PLS1}, which restricts the security gains that arise from spatial beamforming. As shown in Fig. \ref{fig1:(a)}, the conventional secrecy \textit{beam steering} schemes generally utilize the angular dimension to provide security in far-field communications. However, when the E is located in the near-field region, e.g., between the base station (BS) and the legitimate user (U), the eavesdropping channels are highly correlated with legitimate channels in the angular domain, which cannot be efficiently distinguished by the far-field planar-wave channel model. Fortunately, there has been a preliminary study that exploits the distance dimension contained in the spherical-wave channel to secure wireless communications \cite{G.J_NF_PLS}. However, the dedicated secrecy beam focusing strategy for the MIMO network still lacks investigation. Meanwhile, near-field MIMO communications are usually accompanied by extremely large-scale antenna arrays, the fully-digital beamforming structure imposes huge hardware overheads on the network. Therefore, it becomes essential to develop the secrecy beam focusing scheme for MIMO networks with acceptable overheads, which motivates this work.



We propose a near-field secure transmission framework. The secure beam focusing is exploited at the BS to convey the confidential information to a near-field U in the presence of an E located between the U and the BS, as shown in Fig. \ref{fig1:(b)}. The hybrid beamforming architecture is employed at the BS to reduce the radio frequency (RF) chain overhead. A secrecy capacity maximization problem is formulated subject to the analog phase-shift constraints and the baseband digital transmit power budget. A two-stage algorithm is developed to efficiently solve the resulting non-convex problem. Based on the fully-digital beamformers optimized in the first stage, the optimal analog precoders and baseband digital beamformers are alternatingly derived in closed-form expressions. Numerical results demonstrate the convergence of the proposed two-stage algorithm. It also reveals that: 1) the proposed hybrid beamforming scheme can achieve comparable performance to the fully-digital strategy; and 2) the secrecy performance in the near-field systems relies on the distance from the E to the reference point of the U, irrespective of the angle with respect to the BS.

\vspace{-1mm}
\section{System Model and Problem Formulation}
\label{sec:format}

\subsection{System Model}
As shown in Fig. \ref{fig1:(b)}, we consider a near-field MIMO communication system, which consists of a BS, an U and a potential E. The uniform linear array (ULA) is adopted for all the nodes, where the BS is equipped with $M$ antennas, the U is equipped with $M_{\text{U}}$ antennas, and the E is equipped with $M_{\text{E}}$ antennas. The antenna aperture at the BS is assumed to be $D$. The BS operates in the high frequency band (e.g., mmWave or THz), and tries to send the confidential signal to the U in presence of the E. Both U and E are located in near-field region. The distance between the BS and U/E is assumed to be shorter than Rayleigh distance $d_{\text{R}}=\frac{2(D_{1}+D_{2})^{2}}{\lambda}$ ($\lambda$ is the wavelength, $D_{1}$ is the antenna aperture of the BS and $D_{2}$ is the antenna aperture of U). Thus, the transmitted wavefronts follow the spherical propagation. We consider a challenging secure communication scenario, where the E is located in the same direction of the U but closer to the BS than the U. To resist the wiretapping of the E, the BS exploits the beam focusing to enhance the received signal strength at the U while suppressing the information leakage to the E.

In the near-field systems with large number of antennas, the fully-digital beamforming architecture imposes high hardware costs as it requires each antenna to be equipped with a dedicated RF chain. As a result, the hybrid beamforming architecture at the BS is considered \cite{X.Yu_MIMO_Hybrid}. To elaborate, a phase-shift based analog precoder is installed between $M_{\text{R}}$ ($M_{\text{R}}<M$) RF chains and the transmit antenna array, where each output of RF chain is send to all the transmit antennas to form the directional spatial beamformers. Then, $K$ data streams are transmitted to the $M$ transmit antennas via $M_{\text{R}}$ RF chains, which are subject to $K\leq M_{\text{R}}\leq M$. As a result, the transmitted signal at the BS can be expressed as
\begin{equation}
\label{3}
\mathbf{s} = \mathbf{P}\mathbf{W}\mathbf{x},
\end{equation}
where $\mathbf{P}\in\mathbb{C}^{M\times M_{\text{R}}}$ denotes the analog precoding matrix, $\mathbf{W}\in\mathbb{C}^{M_{\text{R}}\times K}$ denotes the digital baseband beamforming matrix, and $\mathbf{x}\in\mathbb{C}^{K\times 1}$ ($\mathbb{E}(\mathbf{x}\mathbf{x}^{H})=\mathbf{I}_{K}$) denotes the data intended for U. Note that the $i$-th row and the $j$-th column element of $\mathbf{P}$ satisfies
\begin{equation}
\label{4}
p_{i,j} \in\mathcal{P}\triangleq\big\{e^{\jmath\vartheta}|\vartheta\in(0,2\pi]\big\},
\end{equation}
where $\vartheta$ represents the phase shift manipulation of $p_{i,j}$. With this process, the received signal at U and E are given by
\begin{align}
\label{5}
\mathbf{y}_{\text{U}}=\mathbf{H}_{\text{B},\text{U}}\mathbf{s}+\mathbf{n}_{\text{U}},\\  \label{6}
\mathbf{y}_{\text{E}}=\mathbf{H}_{\text{B},\text{E}}\mathbf{s}+\mathbf{n}_{\text{E}},
\end{align}
 where $\mathbf{H}_{\text{B},\text{U}}\in\mathbb{C}^{M_{\text{U}}\times M}$ and $\mathbf{H}_{\text{B},\text{E}}\in\mathbb{C}^{M_{\text{E}}\times M}$ denote the equivalent channels from the BS to U and E, $\mathbf{n}_{\text{U}}\sim\mathcal{CN}(0,\sigma^{2}\mathbf{I}_{M_{\text{U}}})$ and $\mathbf{n}_{\text{E}}\sim\mathcal{CN}(0,\sigma^{2}\mathbf{I}_{M_{\text{E}}})$ denote the additive white Gaussian noise (AWGN) at the U and E, respectively. Accordingly, the mutual information between the BS and U/E is given by
 \begin{align}
\label{7}
C_{\text{U}}=\log_{2}\text{det}\left(\mathbf{I}_{M_{\text{U}}}+\sigma^{-2}\mathbf{H}_{\text{B},\text{U}}\mathbf{P}\mathbf{W}
\mathbf{W}^{H}\mathbf{P}^{H}\mathbf{H}_{\text{B},\text{U}}^{H}
\right),\\  \label{8}
C_{\text{E}}=\log_{2}\text{det}\left(\mathbf{I}_{M_{\text{E}}}+\sigma^{-2}\mathbf{H}_{\text{B},\text{E}}\mathbf{P}\mathbf{W}
\mathbf{W}^{H}\mathbf{P}^{H}\mathbf{H}_{\text{B},\text{E}}^{H}
\right).
\end{align}
Following the information-theoretic PLS \cite{M.Bloch_PLS}, the secrecy performance can be characterized by the secrecy capacity, which is defined as the positive difference between the legitimate mutual information and the eavesdropping mutual information, i.e., $C_{\text{s}}=[C_{\text{U}}-C_{\text{E}}]^{+}$, where $[x]^{+}=\max\{x,0\}$ \cite{ZZ_STAR}.

\subsection{Near-Field Channel Model}
For the near-field system, we assume that the coordinate of the midpoint of the BS antenna is $(0,0,0)$. Thus, the $m$-th antenna of the BS can be denoted as $(0,\tilde{m}d,0)$, where $\tilde{m}=m-\frac{M-1}{2}$ and $d$ denotes the antenna pitch. Similarly, the coordinates of the $m_{\text{U}}$-th antenna at the U and the $m_{\text{E}}$-th antenna at the E can be denoted as $(x_{\text{U}},y_{\text{U}}+\tilde{m}_{\text{U}}d,0)$ and $(x_{\text{E}},y_{\text{E}}+\tilde{m}_{\text{E}}d,0)$, where $\tilde{m}_{\text{U}}=m_{\text{U}}-\frac{M_{\text{U}}-1}{2}$ and $\tilde{m}_{\text{E}}=m_{\text{E}}-\frac{M_{\text{E}}-1}{2}$. Accordingly, line-of-sight (LoS) near-field channel between the BS and U can be modeled as  \cite{L.Dai_NF1}
\begin{align}
\label{1}
\mathbf{H}_{\text{B},\text{U}}(d,\theta) = \left[\mathbf{h}_{\text{B},\text{U},1},\cdots,\mathbf{h}_{\text{B},\text{U},M_{\text{U}}}\right]^{T}, 
\end{align}
where $\mathbf{h}_{\text{B},\text{U},m_{\text{U}}}=(1\big/\sqrt{M})\ \big[g_{m_{\text{U}},1}e^{-\jmath \frac{2\pi f}{c} (d_{m_{\text{U}},1}-d_{m_{\text{U}}})},\cdots, \\  g_{m_{\text{U}},M}e^{-\jmath \frac{2\pi f}{c} (d_{m_{\text{U}},M}-d_{m_{\text{U}}})}\big]^{T}$. Note that $|g_{m_{\text{U}},m}|=\frac{c}{4\pi f d_{m_{\text{U}},m}}$ denotes the free-space large-scale path loss between the $m$-th array of the BS and the $m_{\text{U}}$-th antenna of the U, $d_{m_{\text{U}}}$ denotes the reference distance from $(0,0,0)$ to $(x_{\text{U}},y_{\text{U}}+\tilde{m}_{\text{U}}d,0)$, and the distance between the $m$-th array of the BS and the $m_{\text{U}}$-th antenna of the U is given by
\begin{align}
\label{2} \nonumber
d_{m_{\text{U}},m}&=\sqrt{x_{\text{U}}^{2}+[\tilde{m}d-(y_{\text{U}}+\tilde{m}_{\text{U}}d)]^{2}},\\
&=\sqrt{d_{m_{\text{U}}}^{2}+(\tilde{m}d)^{2}-2\tilde{m}d d_{m_{\text{U}}}\sin\theta_{m_{\text{U}}}},
\end{align}
where $\theta_{m_{\text{U}}}$ denotes the azimuth angle of the $m_{\text{U}}$-th antenna of the U with respect to $(0,0,0)$. In the same way, the near-field wiretapping channel $\mathbf{H}_{\text{B},\text{E}}(d,\theta)$ can be obtained. For simplicity, we neglect $(d,\theta)$ in $\mathbf{H}_{\text{B},\text{U}}(d,\theta)$ and $\mathbf{H}_{\text{B},\text{E}}(d,\theta)$ in the following. Note that in contrast to existing works on far-field secure communications \cite{Y.Liu_NOMA_PLS,ZZ_STAR,Q.Shi_MIMO_PLS1}, where the secrecy capacity is significantly degraded by the highly angular correlation between $\mathbf{H}_{\text{B},\text{U}}$ and $\mathbf{H}_{\text{B},\text{E}}$. The spherical-wave channels in the near-field communications contain the extra distance information, which helps to distinguish $\mathbf{H}_{\text{B},\text{U}}$ and $\mathbf{H}_{\text{B},\text{E}}$, and further secures the legitimate transmission.



\vspace{-3mm}
\subsection{Problem Formulation}
In this letter, we aim to maximize the secrecy capacity subject to the analog phase-shift constraints and the transmit power budget of the baseband digital beamformers. The problem formulation is given by
\begin{subequations}
\begin{align}
\label{9a} &\max\limits_{\mathbf{P},\mathbf{W}}\quad C_{\text{s}}\\
\label{9b}&\quad\text{s.t.} \quad \|\mathbf{\tilde{W}}\|^{2}_{\text{F}}\leq P_{\text{max}},\\
\label{9c}&\quad\quad\quad\,\,  p_{i,j}\in\mathcal{P}, 1\leq i\leq M, \quad 1\leq j\leq M_{\text{R}},
\end{align}
\end{subequations}
where $\mathbf{\tilde{W}}\triangleq \mathbf{P}\mathbf{W}$, and $P_{\text{max}}$ denotes the maximal transmit power at the BS.

\section{Secure Beam Focusing Design}
In this section, we investigate the secure beam focusing of the considered near-field system. A two-stage algorithm is developed to optimize the hybrid beamformers. In particular, the block coordinate descent (BCD) approach is employed to design the fully-digital beamformers in the first stage. Then, the analog phase shifts and digital baseband precoders are alternately derived in closed-form expressions.

\subsection{Stage-I: Fully-Digital Beamformer Design}
To provide a performance upper bound for the proposed hybrid architecture, we concentrate on the fully-digital beamformer design in the first stage, where the analog phase-shift constraints is neglected and only the transmit power budget is considered. Accordingly, the problem (9) is reformulated as
\begin{subequations}
\begin{align}
\label{10a} &\max\limits_{\mathbf{W}_{\text{FD}}}\quad C_{\text{s}}\\
\label{10b}&\quad\text{s.t.} \quad \text{Tr}(\mathbf{W}_{\text{FD}}\mathbf{W}_{\text{FD}}^{H})\leq P_{\text{max}}.
\end{align}
\end{subequations}
For notational convenience, we enable $\mathbf{\tilde{H}}_{\text{B},\text{U}}=\sigma^{-1}\mathbf{H}_{\text{B},\text{U}}$ and $\mathbf{\tilde{H}}_{\text{B},\text{E}}=\sigma^{-1}\mathbf{H}_{\text{B},\text{E}}$. Thus, the objective function \eqref{10a} can be expressed as $C_{\text{s}}=\log_{2}\text{det}(\mathbf{I}_{M_{\text{U}}}+\mathbf{\tilde{H}}_{\text{B},\text{U}}\mathbf{W}_{\text{FD}}
\mathbf{W}_{\text{FD}}^{H}\mathbf{\tilde{H}}_{\text{B},\text{U}}^{H})-\log_{2}\text{det}(\mathbf{I}_{M_{\text{E}}}+\mathbf{\tilde{H}}_{\text{B},\text{E}}\mathbf{W}_{\text{FD}}
\mathbf{W}_{\text{FD}}^{H}\mathbf{\tilde{H}}_{\text{B},\text{E}}^{H})$. Note that the problem (10) is challenging to solve due to the intractable Shannon capacity expression in objective function \eqref{10a} and the quadratical power constraint \eqref{10b}. To efficiently tackle this problem, the BCD method is adopted to iteratively solve the problem.
\begin{lemma}\label{Lemma_1} Define a matrix function $\mathbb{F}(\mathbf{U},\mathbf{W})\triangleq(\mathbf{I}-\mathbf{U}^{H}\mathbf{H}\mathbf{W})
    (\mathbf{I}-\mathbf{U}^{H}\mathbf{H}\mathbf{W})^{H}+\mathbf{U}^{H}\mathbf{U}$, the following equalities hold.

    1) The positive definite matrix $\mathbf{V}=(\mathbb{F}(\mathbf{U},\mathbf{W}))^{-1}$ satisfies
    \begin{align}
    \nonumber
    \log\text{det}(\mathbf{I}+\mathbf{H}\mathbf{W}\mathbf{W}^{H}\mathbf{H}^{H}) =  &\max\limits_{\mathbf{V}\succ\mathbf{0},\mathbf{U}}\log\text{det}(\mathbf{V})-\\ \label{11}
    &\text{Tr}(\mathbf{V}
    \mathbb{F}(\mathbf{U},\mathbf{W}))+m,
    \end{align}
    where $\mathbf{U}=(\mathbf{I}+\mathbf{H}\mathbf{W}\mathbf{W}^{H}\mathbf{H}^{H})^{-1}\mathbf{H}\mathbf{W}$.

    2) For any positive definite matrix $\mathbf{E}\in\mathbb{C}^{m\times m}$, we have
    \begin{equation}
    \label{12}
    -\log\text{det}(\mathbf{E}) = \max\limits_{\mathbf{V}\succ\mathbf{0}}\log\text{det}(\mathbf{V})-\text{Tr}(\mathbf{V}\mathbf{E})+m,
    \end{equation}
    where $\mathbf{V}=\mathbf{E}^{-1}$.
\end{lemma}
\begin{IEEEproof}
Please see the proof in \cite[Lemma 4.1]{Q.Shi_MIMO_PLS1}.
\end{IEEEproof}

By substituting $\mathbf{H}=\mathbf{\tilde{H}}_{\text{B},\text{E}}$, $\mathbf{W}=\mathbf{W}_{\text{FD}}$ into \eqref{11} and $\mathbf{E}=\mathbf{I}_{M_\text{E}}+\mathbf{\tilde{H}}_{\text{B},\text{E}}\mathbf{W}_{\text{FD}}\mathbf{W}_{\text{FD}}^{H}\mathbf{\tilde{H}}_{\text{B},\text{E}}^{H}$ into \eqref{12}, the problem (10) can be reformulated as
\begin{subequations}
\begin{align}
\nonumber
 &\max\limits_{\mathbf{W}_{\text{FD}},\mathbf{V}_{\text{U}}\succ\mathbf{0},\mathbf{V}_{\text{E}}\succ\mathbf{0},\mathbf{U}}\ \log\text{det}(\mathbf{V}_{\text{U}})-\text{Tr}(\mathbf{V}_{\text{U}}
    \mathbb{F}_{\text{U}}(\mathbf{U},\mathbf{W}_{\text{FD}}))+K\\ \label{13a}
    &+\log\text{det}(\mathbf{V}_{\text{E}})-\text{Tr}(\mathbf{V}_{\text{E}}
    (\mathbf{I}_{M_\text{E}}+\mathbf{\tilde{H}}_{\text{B},\text{E}}\mathbf{W}_{\text{FD}}\mathbf{W}_{\text{FD}}^{H}\mathbf{\tilde{H}}_{\text{B},\text{E}}^{H}))+M_\text{E}\\
\label{13b}&\qquad\qquad\qquad\text{s.t.} \quad \text{Tr}(\mathbf{W}_{\text{FD}}\mathbf{W}_{\text{FD}}^{H})\leq P_{\text{max}},
\end{align}
\end{subequations}
where $\{\mathbf{U},\mathbf{V}_{\text{U}},\mathbf{V}_{\text{E}}\}$ are the introduced auxiliary variables, and $\mathbb{F}_{\text{U}}(\mathbf{U},\mathbf{W}_{\text{FD}})\triangleq(\mathbf{I}-\mathbf{U}^{H}\mathbf{\tilde{H}}_{\text{B},\text{U}}\mathbf{W}_{\text{FD}})
    (\mathbf{I}-\mathbf{U}^{H}\mathbf{\tilde{H}}_{\text{B},\text{U}}\mathbf{W}_{\text{FD}})^{H}+\mathbf{U}^{H}\mathbf{U}$. In the following, we solve the problem (13) iteratively by employing the BCD approach. To elaborate, the optimization variables are divided into three blocks, i.e., $\{\mathbf{U}\}$, $\{\mathbf{V}_{\text{U}},\mathbf{V}_{\text{E}}\}$ and $\{\mathbf{W}_{\text{FD}}\}$. In each iteration, we optimize the optimization variables in one block while remaining the other blocks constant.
\subsubsection{Subproblem with respect to $\{\mathbf{U}\}$} By fixing $\{\mathbf{V}_{\text{U}},\mathbf{V}_{\text{E}}\}$ and $\{\mathbf{W}_{\text{FD}}\}$, the problem (13) is reduced to $\min\limits_{\mathbf{U}}\  \text{Tr}(\mathbf{V}_{\text{U}} \mathbb{F}_{\text{U}}(\mathbf{U},\mathbf{W}_{\text{FD}}))$. According to Lemma \ref{Lemma_1}, the optimal solution of $\mathbf{U}$ can be derived in the following expression.
\begin{equation}
    \label{14}
    \mathbf{U}^{*}=(\mathbf{I}_{M_{\text{U}}}+\mathbf{\tilde{H}}_{\text{B},\text{U}}\mathbf{W}_{\text{FD}}\mathbf{W}_{\text{FD}}^{H}
    \mathbf{\tilde{H}}_{\text{B},\text{U}}^{H})^{-1}\mathbf{\tilde{H}}_{\text{B},\text{U}}\mathbf{W}_{\text{FD}}.
\end{equation}

\subsubsection{Subproblem with respect to $\{\mathbf{V}_{\text{U}},\mathbf{V}_{\text{E}}\}$} With fixed $\{\mathbf{U}\}$ and $\{\mathbf{W}_{\text{FD}}\}$, the problem (13) is reduced to two separate subproblems, i.e.,  $\max\limits_{\mathbf{V}_{\text{U}} }\  \log\text{det}(\mathbf{V}_{\text{U}})-\text{Tr}(\mathbf{V}_{\text{U}}\succeq\mathbf{0} \mathbb{F}_{\text{U}}(\mathbf{U},\mathbf{W}_{\text{FD}}))$ and $\max\limits_{\mathbf{V}_{\text{E}}\succeq\mathbf{0} }\  \log\text{det}(\mathbf{V}_{\text{E}})-\text{Tr}(\mathbf{V}_{\text{E}}
    (\mathbf{I}_{M_\text{E}}+\mathbf{\tilde{H}}_{\text{B},\text{E}}\mathbf{W}_{\text{FD}}\mathbf{W}_{\text{FD}}^{H}\mathbf{\tilde{H}}_{\text{B},\text{E}}^{H}))$. With condition for the equal sign to hold, we can derive the optimal solution of $\{\mathbf{V}_{\text{U}},\mathbf{V}_{\text{E}}\}$, which is given by
    \begin{gather}
    \label{15}
    \mathbf{V}_{\text{U}}^{*}\!=\!\left((\mathbf{I}\!-\!\mathbf{U}^{H}\mathbf{\tilde{H}}_{\text{B},\text{U}}\mathbf{W}_{\text{FD}})
    (\mathbf{I}\!-\!\mathbf{U}^{H}\mathbf{\tilde{H}}_{\text{B},\text{U}}\mathbf{W}_{\text{FD}})^{H}\!\!+\!\mathbf{U}^{H}\mathbf{U}\right)^{\!-1},\\
    \label{16}
    \mathbf{V}_{\text{E}}^{*}=\left(\mathbf{I}_{M_\text{E}}+\mathbf{\tilde{H}}_{\text{B},\text{E}}
    \mathbf{W}_{\text{FD}}\mathbf{W}_{\text{FD}}^{H}\mathbf{\tilde{H}}_{\text{B},\text{E}}^{H}\right)^{-1}.
    \end{gather}

\subsubsection{Subproblem with respect to $\{\mathbf{W}_{\text{FD}}\}$} Solving problem (13) for $\mathbf{W}_{\text{FD}}$ with given $\{\mathbf{U}\}$ and $\{\mathbf{V}_{\text{U}},\mathbf{V}_{\text{E}}\}$ is equivalent to the following subproblem.
\begin{subequations}
\begin{align}
\nonumber
 \min\limits_{\mathbf{W}_{\text{FD}}}\quad &\text{Tr}(\mathbf{V}_{\text{U}}
    \mathbb{F}_{\text{U}}(\mathbf{U},\mathbf{W}_{\text{FD}}))+\\ \label{17a}
    &\text{Tr}(\mathbf{V}_{\text{E}}
    (\mathbf{I}_{M_\text{E}}+\mathbf{\tilde{H}}_{\text{B},\text{E}}\mathbf{W}_{\text{FD}}\mathbf{W}_{\text{FD}}^{H}\mathbf{\tilde{H}}_{\text{B},\text{E}}^{H}))\\
\label{17b}\quad\text{s.t.} \quad &\text{Tr}(\mathbf{W}_{\text{FD}}\mathbf{W}_{\text{FD}}^{H})\leq P_{\text{max}},
\end{align}
\end{subequations}
Note problem (17) is a convex a second order cone programming (SOCP) program, which can be optimally solved. However, the near-field systems are usually accompanied by extremely large-scale antenna arrays. It indicates that to directly solve problem (17) possesses a high computational complexity, which is not applicable in practice. Since problem (17) is convex and satisfies Slater's condition, the strong duality holds between the original problem and the dual problem \cite{S.Boyd}. Thus, we can obtain the optimal solution of problem (17) by solving its dual problem, where the Lagrangian function with respect to $\mathbf{W}_{\text{FD}}$ is given by
\begin{align}
    \nonumber
    &\mathcal{L}(\mathbf{W}_{\text{FD}},\mu)=\text{Tr}(\mathbf{W}_{\text{FD}}\mathbf{\tilde{H}}_{\text{B},\text{U}}^{H}
    \mathbf{U}\mathbf{V}_{\text{U}}\mathbf{U}^{H}\mathbf{\tilde{H}}_{\text{B},\text{U}}\mathbf{W}_{\text{FD}}^{H})-\\ \nonumber
    &\ \ \text{Tr}(\mathbf{V}_{\text{U}}\mathbf{U}^{H}\mathbf{\tilde{H}}_{\text{B},\text{U}}\mathbf{W}_{\text{FD}})-
    \text{Tr}(\mathbf{V}_{\text{U}}\mathbf{W}_{\text{FD}}^{H}\mathbf{\tilde{H}}_{\text{B},\text{U}}^{H}\mathbf{U})+\\ \label{18}
    &\ \
    \text{Tr}(\mathbf{W}_{\text{FD}}\mathbf{\tilde{H}}_{\text{B},\text{E}}^{H}
    \mathbf{V}_{\text{E}}\mathbf{\tilde{H}}_{\text{B},\text{E}}\mathbf{W}_{\text{FD}}^{H})+\mu(\text{Tr}(
    \mathbf{W}_{\text{FD}}\mathbf{W}_{\text{FD}}^{H})-P_{\text{max}}),
\end{align}
where $\mu\geq 0$ is the Lagrangian multiplier. By defining $f(\mu)=\min\limits_{\mathbf{W}_{\text{FD}}}\mathcal{L}(\mathbf{W}_{\text{FD}},\mu)$, the dual problem is given by
\begin{subequations}
\begin{align}
\label{19a}
 \max\limits_{\mu}\quad &f(\mu)\\
\label{19b}\quad\text{s.t.} \quad &\mu\geq 0.
\end{align}
\end{subequations}
Note that the optimal solution of $\mathbf{W}_{\text{FD}}$ under any given $\mu>0$ can be derived by adopting the first-order optimality condition, i.e.,
\begin{equation}
    \label{20}
    \mathbf{W}_{\text{FD}}^{*}(\mu)=\mathbf{E}(\mu\mathbf{I}_{M}+\mathbf{D})^{-1}\mathbf{E}^{H}
    \mathbf{\tilde{H}}_{\text{B},\text{U}}^{H}\mathbf{U}\mathbf{\tilde{H}}_{\text{B},\text{U}},
\end{equation}
where $\mathbf{E}\mathbf{D}\mathbf{E}^{H}$ is the result of the eigen-decomposition of $\mathbf{\tilde{H}}_{\text{B},\text{U}}^{H}
    \mathbf{U}\mathbf{V}_{\text{U}}\mathbf{U}^{H}\mathbf{\tilde{H}}_{\text{B},\text{U}}+
    \mathbf{\tilde{H}}_{\text{B},\text{E}}^{H}
    \mathbf{V}_{\text{U}}\mathbf{\tilde{H}}_{\text{B},\text{E}}$. With the above derivation, the optimal $\mathbf{W}_{\text{FD}}(\mu)$ can be obtained via the one-dimensional search for $\mu$, which can be efficiently dealt with by the Bisection method \cite{S.Boyd}. Then, by iteratively updating $\{\mathbf{V}_{\text{U}},\mathbf{V}_{\text{E}}\}$ and $\{\mathbf{W}_{\text{FD}}\}$, we can obtain the fully-digital beamformers of the considered network.
\begin{algorithm}[t]
    \caption{Two-stage algorithm.}
    \label{two-phase}
    \begin{algorithmic}[1]
        \STATE{Initialize initial $\mathbf{P}$ and $\mathbf{W}_{\text{FD}}$ with $n=1$ and $m=1$. Set the convergence accuracy $\epsilon_{1}$ and $\epsilon_{2}$.}
        \STATE{\textbf{BCD repeat}}
        \STATE{ \quad update $\mathbf{U}^{n}$ according to \eqref{14}.}
        \STATE{ \quad update $\{\mathbf{V}_{\text{U}}^{n},\mathbf{V}_{\text{E}}^{n}\}$ according to \eqref{15} and \eqref{16}.}
        \STATE{ \quad update $\mathbf{W}_{\text{FD}}^{n}$ by carrying out \textbf{Algorithm \ref{bisection}}.}
        \STATE{ \quad set $n=n+1$.}
        \STATE{ \textbf{until} the $|C_{\text{s}}(\mathbf{W}_{\text{FD}}^{n})-C_{\text{s}}(\mathbf{W}_{\text{FD}}^{n-1})|\leq\epsilon_{1}$.}
        \STATE{\textbf{AO repeat}}
        \STATE{ \quad update $\mathbf{W}^{m}$ according to \eqref{22}.}
        \STATE{ \quad iteratively update $p(i,j)^{m}$ according to \eqref{25}.}
        \STATE{ \quad set $m=m+1$.}
        \STATE{ \textbf{until} the $|C_{\text{s}}(\mathbf{P}^{m}\mathbf{W}^{m})-C_{\text{s}}(\mathbf{P}^{m+1}\mathbf{W}^{m+1})|\leq\epsilon_{2}$.}
    \end{algorithmic}
\end{algorithm}

\vspace{-3mm}
\subsection{Stage-II: Hybrid Beamformer Design}
In this subsection, we focus on the design of the hybrid beamformers. To approximately maximize the secrecy mutual information between the BS and U \cite{O.El_mmWave_MIMO}, we project the optimized $\mathbf{W}_{\text{FD}}$ to the set of hybrid beamformers to obtain the near-optimal analog phase shifters and baseband precoders. The hybrid beamformer design problem is given by
\begin{subequations}
\begin{align}
\label{21a}
 \min\limits_{\mathbf{P},\mathbf{W}}\quad &\|\mathbf{W}_{\text{FD}}-\mathbf{P}\mathbf{W}\|^{2}_{\text{F}}\\
\label{21b}\quad\text{s.t.} \quad & p_{i,j}\in\mathcal{P}, 1\leq i\leq M, \quad 1\leq j\leq M_{\text{R}}.
\end{align}
\end{subequations}
Notably, problem (21) is a highly coupled quadratic problem, so we consider adopting the alternating optimization (AO) framework to iteratively optimize the digital baseband precoder and the analog phase shifters.
\subsubsection{Digital Baseband Precoder Design} With the fixed $\mathbf{P}$, the problem (21) is reduced to $\min\limits_{\mathbf{W}}\  \|\mathbf{W}_{\text{FD}}-\mathbf{P}\mathbf{W}\|^{2}_{\text{F}}$, which can be optimally solved by adopting the first-order optimality condition. As such, the optimal $\mathbf{W}$ is given by
\begin{equation}
    \label{22}
    \mathbf{W}^{*}=(\mathbf{P}^{H}\mathbf{P})^{-1}\mathbf{P}^{H}\mathbf{W}_{\text{FD}}.
\end{equation}
\subsubsection{Analog Phase Shifter Design} With the fixed $\mathbf{W}$, the problem (21) can be reduced as
\begin{subequations}
\begin{align}
\label{23a}
 \min\limits_{\mathbf{P}}\quad &\text{Tr}(\mathbf{P}^{H}\mathbf{P}\mathbf{X})-2\Re(\text{Tr}(\mathbf{P}\mathbf{Y}))\\
\label{23b}\quad\text{s.t.} \quad & p_{i,j}\in\mathcal{P}, 1\leq i\leq M, \quad 1\leq j\leq M_{\text{R}},
\end{align}
\end{subequations}
where $\mathbf{X}=\mathbf{W}\mathbf{W}^{H}$ and $\mathbf{Y}=\mathbf{W}_{\text{FD}}\mathbf{W}^{H}$. Since the variable $p_{i,j}$ are separable in the unit-modulus constraint \eqref{23b}, the problem (23) can be efficiently tackled by the BCD method, which iteratively optimizes each entry of $\mathbf{P}$ while fixing the remaining elements. Consequently, the subproblem with respect to $p_{i,j}$ is given by
\begin{subequations}
\begin{align}
\label{24a}
 \max\limits_{|p_{i,j}|=1}\quad &\Re(z_{i,j}p_{i,j}),
\end{align}
\end{subequations}
where $z_{i,j}$ is a complex coefficient determined by the elements of $\mathbf{P}$ except for $p_{i,j}$. Under the unit-modulus constraint, the optimal $p_{i,j}$ can be derived as follows.
\begin{equation}
    \label{25}
    p_{i,j}^{*}=\frac{z_{i,j}}{|z_{i,j}|},
\end{equation}
where $z_{i,j}=\mathbf{Y}_{[j,j]}-(\mathbf{\tilde{X}}_{[i,j]}-p_{i,j}\mathbf{X}_{[j,j]})$ and $\mathbf{\tilde{X}}=\mathbf{P}\mathbf{X}$. Afterwards, by alternatingly updating $\mathbf{W}$ and $p_{i,j}$, the digital baseband precoders and analog phase shifters can be determined.

\begin{algorithm}[t]
    \caption{Bisection algorithm.}
    \label{bisection}
    \begin{algorithmic}[1]
        \STATE{Initialize initial $\mu_{\text{lower}}$ and $\mu_{\text{upper}}$. Set a convergence accuracy $\epsilon_{3}$.}
        \REPEAT
        \STATE{ $\mu=\frac{\mu_{\text{lower}}+\mu_{\text{upper}}}{2}$.}
        \STATE{ update $\mathbf{W}_{\text{FD}}$ according to \eqref{20}.}
        \STATE{ \textbf{if} $\text{Tr}(\mathbf{W}_{\text{FD}}\mathbf{W}_{\text{FD}}^{H})\leq P_{\text{max}}$}
        \STATE{  \quad $\mu_{\text{lower}}=\mu$,}
        \STATE{ \textbf{else}}
        \STATE{  \quad $\mu_{\text{upper}}=\mu$,}
        \STATE{ \textbf{end}}
        \UNTIL{ the $|\text{Tr}(\mathbf{W}_{\text{FD}}\mathbf{W}_{\text{FD}}^{H})- P_{\text{max}}|\leq\epsilon_{3}$.}
    \end{algorithmic}
\end{algorithm}

\vspace{-3mm}
\subsection{Overall Algorithm}
The proposed two-stage algorithm is summarized in \textbf{Algorithm \ref{two-phase}}. For the BCD loop in \textbf{Algorithm \ref{two-phase}}, since the optimal solutions $\{\mathbf{U}\}$ and $\{\mathbf{V}_{\text{U}},\mathbf{V}_{\text{E}}\}$ and the Karush-Kuhn-Tucker (KKT) point solution $\{\mathbf{W}_{\text{FD}}\}$ are guaranteed in steps 3, 4 and 5, we readily have the following inequality
\begin{align}
    \nonumber
    &C_{\text{s}}(\mathbf{U}^{n},\mathbf{V}_{\text{U}}^{n},\mathbf{V}_{\text{E}}^{n},\mathbf{W}_{\text{FD}}^{n})\geq
    C_{\text{s}}(\mathbf{U}^{n+1},\mathbf{V}_{\text{U}}^{n},\mathbf{V}_{\text{E}}^{n},\mathbf{W}_{\text{FD}}^{n})\geq \\ \nonumber
    &\qquad C_{\text{s}}(\mathbf{U}^{n+1},\mathbf{V}_{\text{U}}^{n+1},\mathbf{V}_{\text{E}}^{n+1},\mathbf{W}_{\text{FD}}^{n})\geq \\ \label{26}
    &\qquad C_{\text{s}}(\mathbf{U}^{n+1},\mathbf{V}_{\text{U}}^{n+1},\mathbf{V}_{\text{E}}^{n+1},\mathbf{W}_{\text{FD}}^{n+1}),
\end{align}
which proves the monotonic convergence of the generated sequence $\{C_{\text{s}}^{n},\cdots,C_{\text{s}}^{n+m},\cdots\}$ with $C_{\text{s}}^{n}= C_{\text{s}}(\mathbf{U}^{n},\mathbf{V}_{\text{U}}^{n},\mathbf{V}_{\text{E}}^{n},\\ \mathbf{W}_{\text{FD}}^{n})$. Furthermore, by checking the KKT conditions, it is readily know that the accumulation point $\bar{C}_{\text{s}}$ of the sequence $\{C_{\text{s}}^{n},\cdots,C_{\text{s}}^{n+m},\cdots\}$ is the KKT solution of the original problem \cite[Proposition 4.2]{Q.Shi_MIMO_PLS1}. In the same way, we can prove that the AO alternating iteration converges to at least the stationary point solution of the problem (21).

Since all the subproblems are solved by the closed-form solutions, so the proposed two-stage algorithm is complexity-efficient. The main complexity of the proposed two-stage algorithm relies on the eigen-decomposition operation and inverse matrix operation, the whole complexity is given by $\mathcal{O}\Big(l_{1}(K^{3}+M_{\text{R}}^{3}+(l_{\text{B}}+1)M^{3})+l_{2}K^{3}\Big)$ \cite{J.Nocedal}, where $l_{1}$, $l_{\text{B}}$ and $l_{2}$ denote the number of iterations of the BCD loop, the Bisection algorithm, and the AO loop.

\vspace{-2mm}
\section{Numerical Results}
This section provides the numerical results to validate the effectiveness of the proposed scheme. The linear topology is considered for the simulations, where the midpoint of the BS antenna is located in (0,0,0) meter (m), while midpoints of the antennas of U and E are respectively located 15 m and 5 m from the coordinate (0,0,0) m with the azimuth angle of $45^{\circ}$. All the ULA are positioned along the y-axis. Unless otherwise specified, the default parameters are set as $f=28$ GHz, $d=\frac{\lambda}{2}$, $M = 256$, $M_\text{U} = 8$, $M_\text{E} = 8$, $M_\text{R}=4$, $K = 2$, $\sigma^{2}=-105$ dBm, $\epsilon_{1}=10^{-4}$ and $\epsilon_{2}=\epsilon_{3}=10^{-6}$. The numerical results are averaged from 100 independent Monte-Carlo experiments.

\begin{figure}[!t]
 \centering
 \subfigure[Convergence of the proposed two-stage algorithm with $P_{\text{max}}=-10$ dBm.]{
  \includegraphics[scale = 0.27]{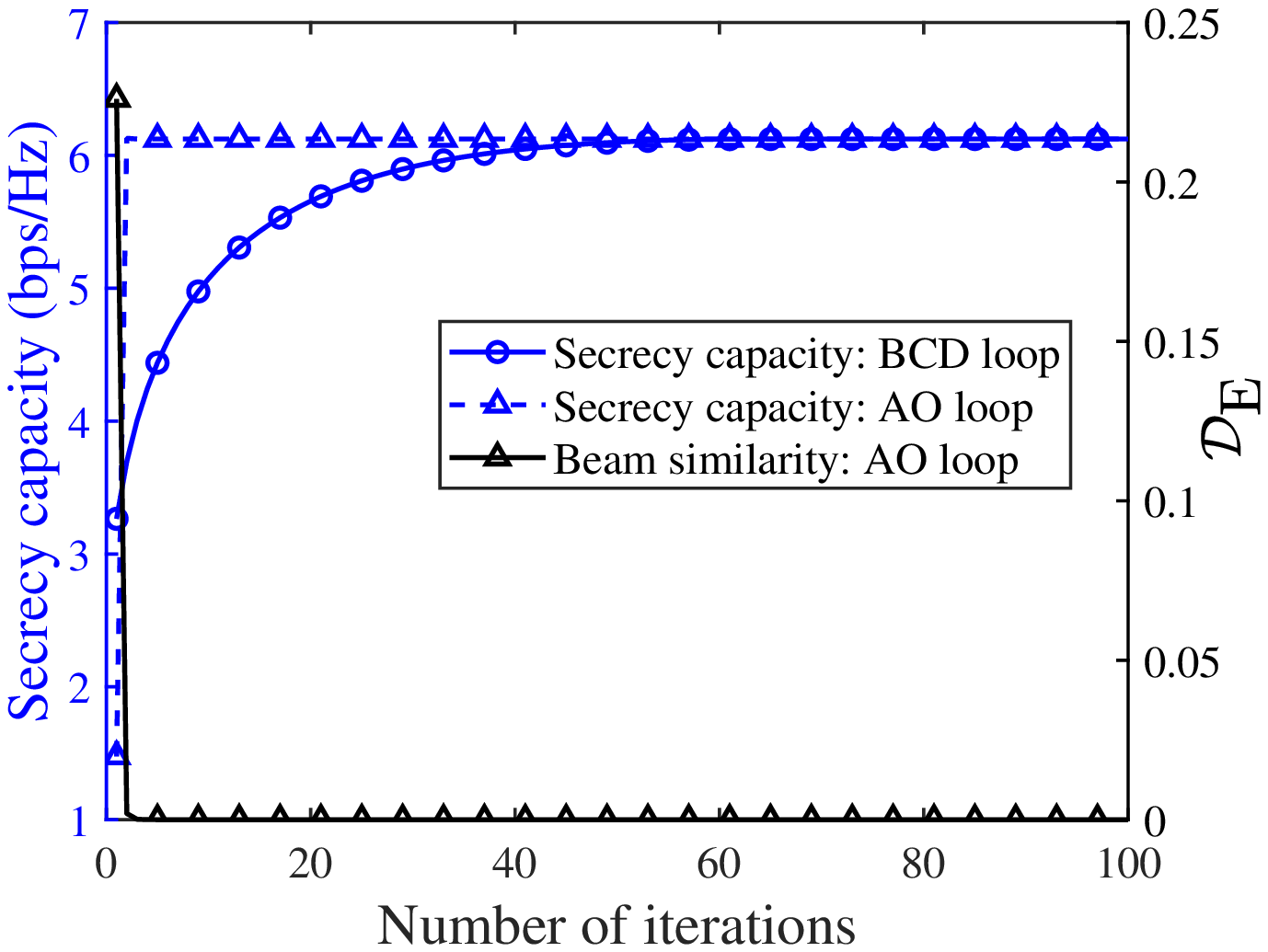}
   \label{fig2}
 }
\subfigure[Secrecy performance comparison versus $P_{\text{max}}$ for different simulation parameters.]{
 \includegraphics[scale = 0.27]{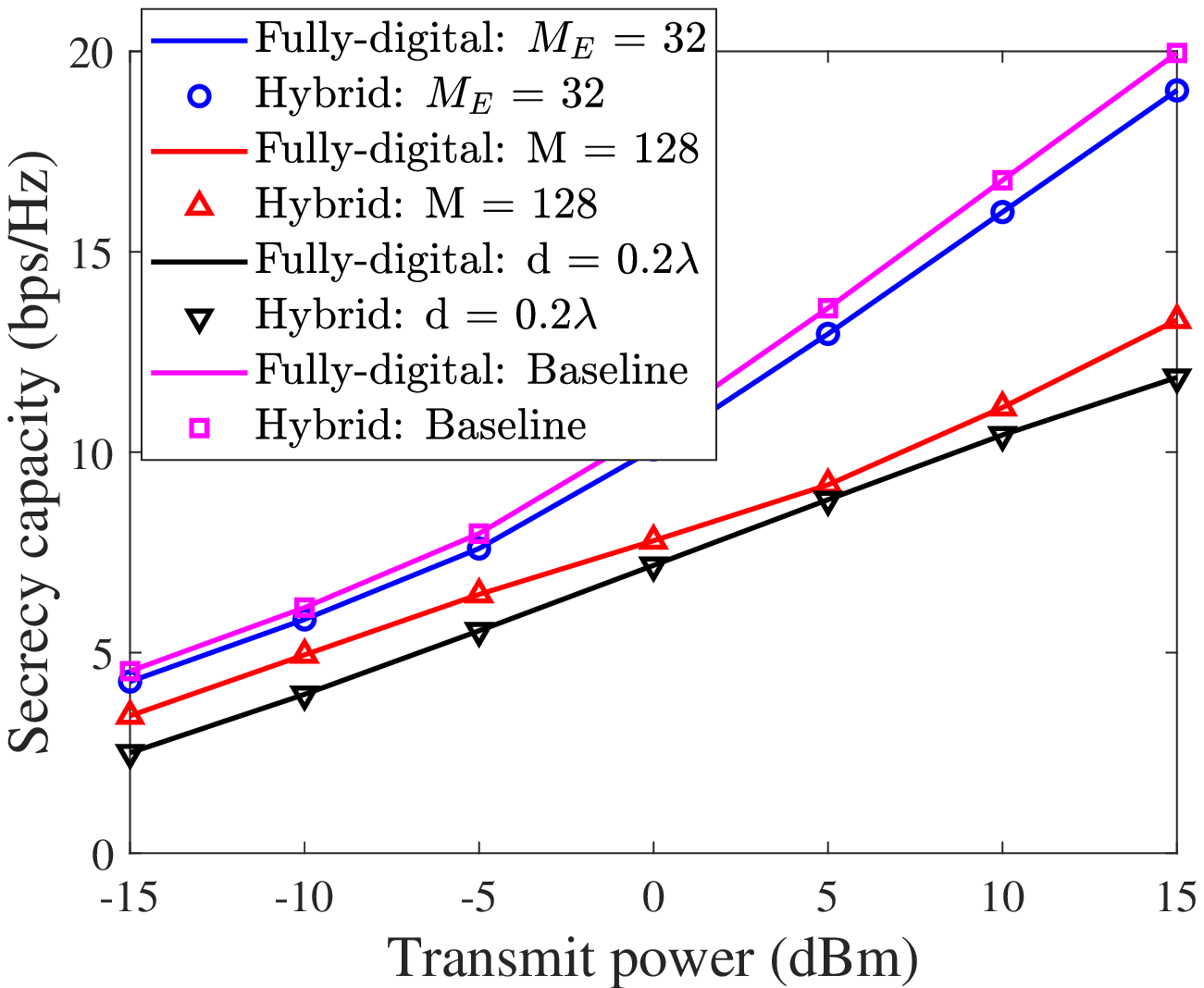}
 \label{fig3}
}
\caption{Algorithm performance evaluation.}
\label{fig:2}
\end{figure}


Fig. \ref{fig2} depicts the convergence performance of \textbf{Algorithm \ref{two-phase}}, where the beam similarity in objective function \eqref{21a} is represented as $D_{\text{E}}\triangleq\|\mathbf{W}_{\text{FD}}-\mathbf{P}\mathbf{W}\|^{2}_{\text{F}}$. As can be seen, both the BCD loop and the AO loop in the proposed two-stage algorithm can monotonically converge the stationary point solutions within the finite iterations, which demonstrates the effectiveness of the proposed scheme. It can be also observed that the optimized hybrid beamformers can achieve the comparable performance to the fully-digital beamformers. This result can be expected since for each subproblem, the optimal analog phase shifters and baseband digital precoders are alternatingly derived in closed-form expressions in the AO loop.

Fig. \ref{fig3} illustrates the secrecy performance of the proposed algorithm, where the baseline scheme is under the default parameters. It is observed that decreasing transmit antennas or increasing E's antennas degrades the secrecy performance of the system. It is because that decreasing the transmit antennas reduces the spatial degrees-of-freedom (DoFs) that the BS can exploit, and meanwhile, increasing the antennas of E enhance the E's reception ability. Both of them narrow the gap between the legitimate channel capacity and the eavesdropping channel capacity, thus deteriorating the secrecy performance. We can also see that the secrecy capacity increases with the increasing $d$. It is due to the fact that increasing $d$ improves the antenna aperture, which leads to a large near-field region and enhance the angular/distance resolution of the beam focusing.

\begin{figure}[!t]
 \centering
 \subfigure[Secrecy performance versus the location of E, where $P_{\text{max}}=-15$ dBm.]{
  \includegraphics[scale = 0.27]{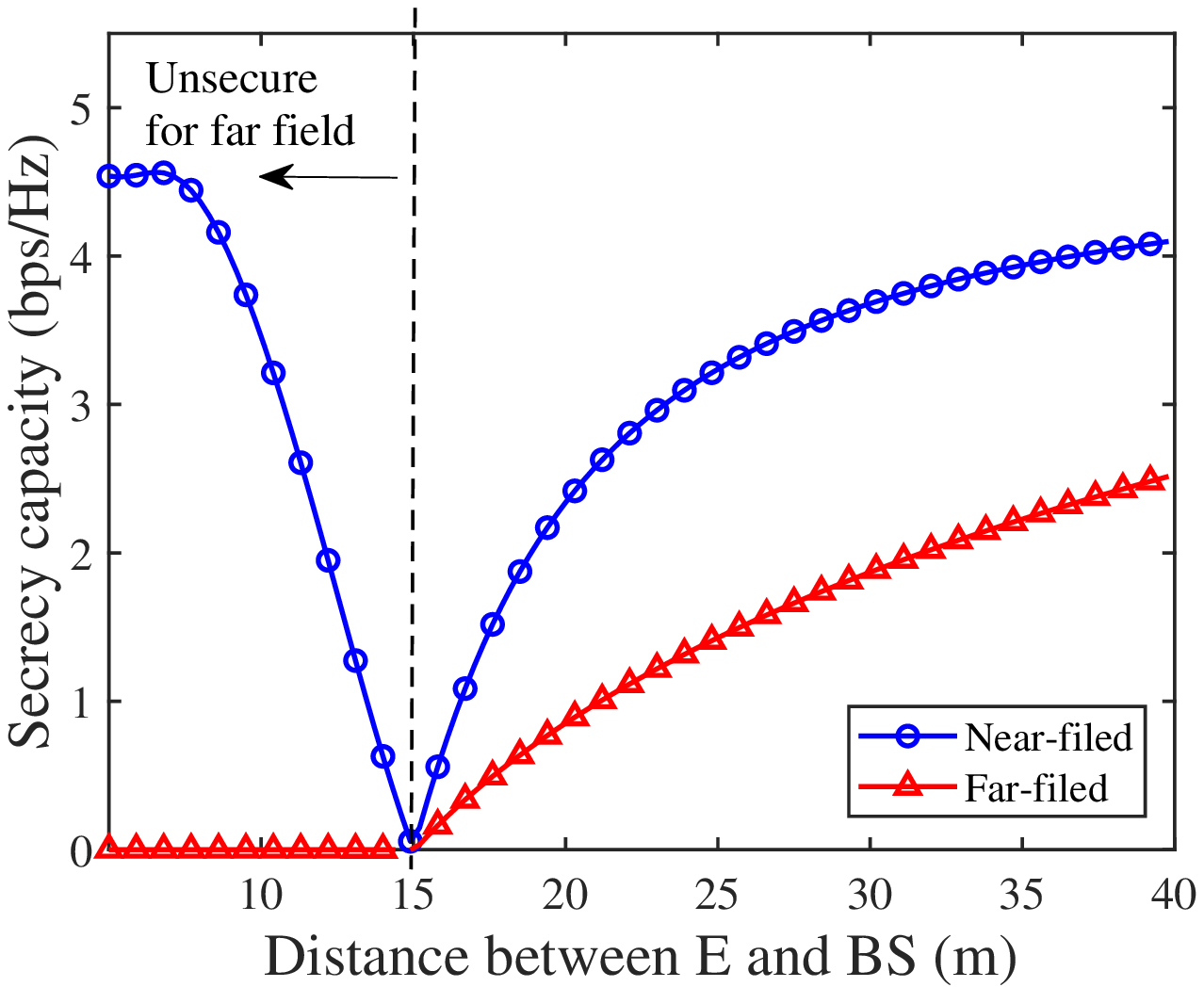}
   \label{fig4}
 }
\subfigure[Normalized signal power spectrum, where $P_{\text{max}}=-15$ dBm, U is located 20 m from the BS, and E is located 10 m from the BS.]{
 \includegraphics[scale = 0.27]{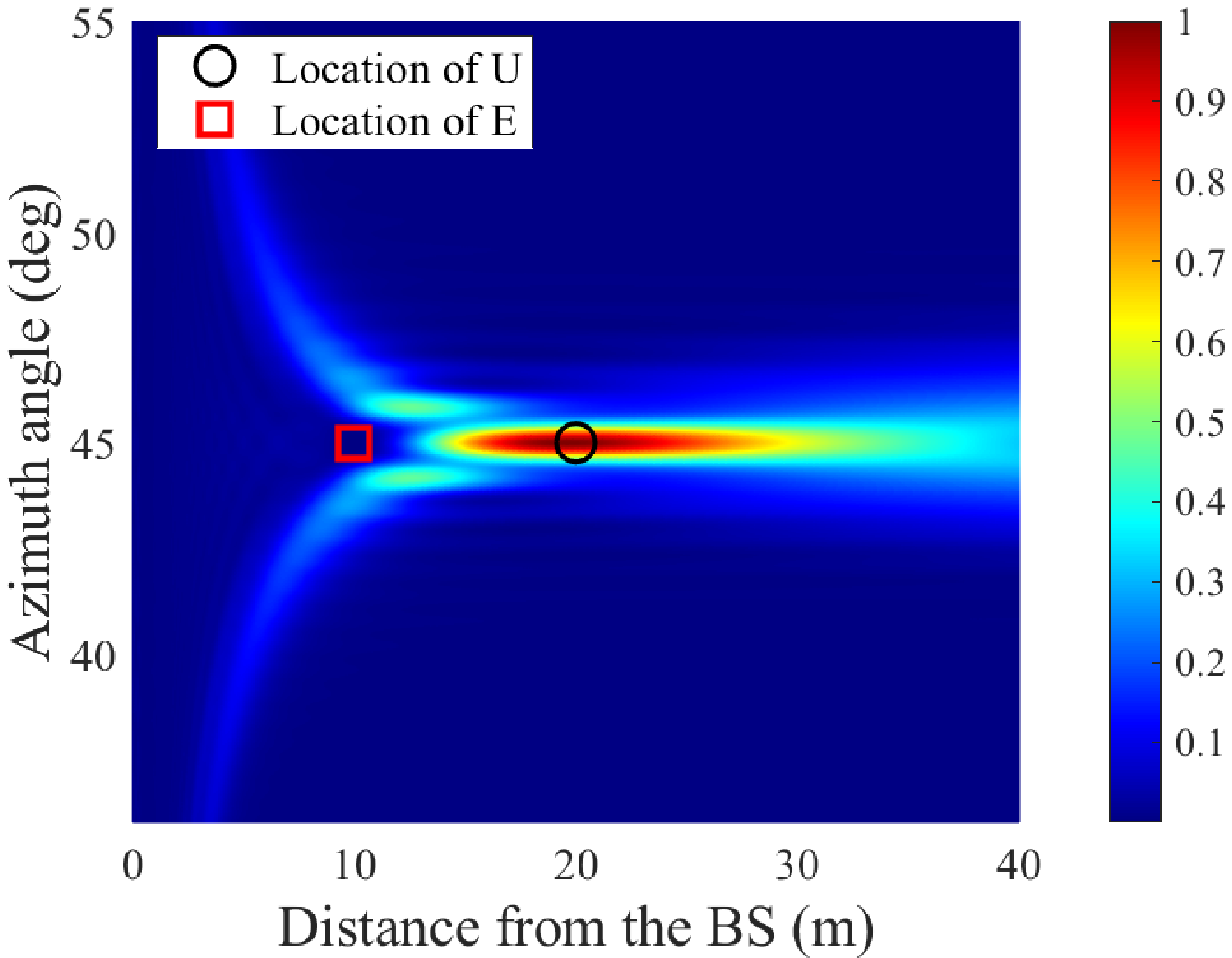}
 \label{fig5}
}\caption{Security gains with near-field beam focusing.}

\label{fig:3}
\end{figure}

In Fig. \ref{fig4}, we present the secrecy capacity versus the location of E. It can be seen that in the far-field communication, when the E is positioned in the same direction as the U, perfectly secure transmission only occurs when the eavesdropping links suffers worse channel conditions than the legitimate links. However, in the near-field communication, the perfectly secure transmission is always guaranteed, except when the E has the same position as the U. This is because that in the far-field communication, the secrecy performance is mainly dependent on the angular disparity between the U and the E with respect to the BS as the reference point. While in the near-field communication, the secrecy performance mainly relies on the distance disparity of the E with respect to the reference point of the U.

To further illustrate the impact of beam focusing in near-field communications, Fig. \ref{fig5} plots the normalized signal power spectrum over the free-space location. As can be observed, the optimized beamformers can directionally enhance the signal power at the direction of $45^{\circ}$. Meanwhile, we can also see that at a distance of 10 m, i.e. at the position of E, the signal is fully suppressed, while at a distance of 20 m, the signal power is significantly strengthened. This result demonstrates that the proposed secure beam focusing scheme can precisely enhance the signal strength at a specific point of free space without significant energy/information leakage on the incident pathes.

\vspace{-3mm}
\section{Conclusion}
A novel secure near-field framework was proposed. A two-stage algorithm was developed to maximize the secrecy capacity of the U via jointly optimizing unit-modulus phase shifters and baseband digital beamformers. Numerical results were present to unveil that the secrecy performance of near-field communications is primarily relevant to the relative distance of the E with respect to the U.


\vspace{-3mm}
\begin{thebibliography}{99}

\bibitem{W.Saad_6G}
W. Saad, M. Bennis, and M. Chen, ``A vision of 6G wireless systems: Applications, trends, technologies, and open research problems,'' \textit{IEEE
Netw.}, vol. 34, no. 3, pp. 134--142, May. 2020.

\bibitem{H.Zhang_NF}
H. Zhang, N. Shlezinger, et al, ``Beam focusing for near-field multiuser MIMO communications,'' \textit{IEEE Trans. Wireless Commun.}, vol. 21, no. 9, pp. 7476--7490, Sep. 2022.

\bibitem{C.Huang_NF2}
L. Wei, C. Huang, et al, ``Tri-polarized holographic MIMO surface in near-field: Channel modeling and precoding design,'' [Online]. Available: https://arxiv.org/pdf/2211.03479

\bibitem{C.Huang_NF1}
X. Gan, C. Huang, Z. Yang, C. Zhong, and Z. Zhang, ``Near-field localization for holographic RIS assisted mmWave systems,'' \textit{IEEE Commun. Lett.}, vol. 27, no. 1, pp. 140--144, Jan. 2023.

\bibitem{J.Xu_STAR-RIS}
J. Xu, Y. Liu, X. Mu, and O. A. Dobre, ``STAR-RISs: Simultaneous transmitting and reflecting reconfigurable intelligent surfaces,''
\textit{IEEE Commun. Lett.}, vol. 25, no. 9, pp. 3134--3138, Sep. 2021.


\bibitem{L.Dai_NF1}
M. Cui, L. Dai, R. Schober, and L. Hanzo, ``Near-field wideband beamforming for extremely large antenna arrays,'' [Online]. Available: https://arxiv.org/pdf/2109.10054v1

\bibitem{M.Bloch_PLS}
M. Bloch and J. Barros, \textit{Physical-Layer Security: From Information Theory to Security Engineering.} Cambridge, U.K.: Cambridge Univ. Press, 2011.

\bibitem{Y.Liu_NOMA_PLS}
Y. Liu, Z. Qin, M. Elkashlan, Y. Gao, and L. Hanzo, ``Enhancing the physical layer security of non-orthogonal multiple access in large-scale networks,'' \textit{IEEE Trans. Wireless Commun.}, vol. 16, no. 3, pp. 1656--1672, Mar. 2017.


\bibitem{ZZ_STAR}
Z. Zhang, J. Chen, Y. Liu, Q. Wu, B. He, and L. Yang, ``On the secrecy design of STAR-RIS assisted uplink NOMA networks,'' \textit{IEEE Trans. Wireless Commun.}, vol. 21, no. 12, pp. 11207--11221, Dec. 2022.


\bibitem{Q.Shi_MIMO_PLS1}
Q. Shi, W. Xu, J. Wu, E. Song, and Y. Wang, ``Secure beamforming for MIMO broadcasting with wireless information and power transfer,'' \textit{IEEE Trans. Wireless Commun.}, vol. 14, no. 5, pp. 2841--2853, May. 2015.

\bibitem{G.J_NF_PLS}
G. J. Anaya-Lopez, J. P. Gonzalez-Coma, et al, ``Spatial degrees of freedom for physical layer security in XL-MIMO,'' in \textit{Proc. IEEE 95th Veh. Technol. Conf. (VTC-Spring)}, Helsinki, Finland, Jun. 2022, pp. 1-5

\bibitem{X.Yu_MIMO_Hybrid}
X. Yu, J. -C. Shen, et al, ``Alternating minimization algorithms for hybrid precoding in millimeter wave MIMO systems,'' \textit{IEEE J. Sel. Top. Signal Process.}, vol. 10, no. 3, pp. 485--500, Apr. 2016.


\bibitem{S.Boyd}
S. Boyd and L. Vandenberghe, \textit{Convex Optimization}. Cambridge, U.K.:
Cambridge Univ. Press, 2004.

\bibitem{J.Nocedal}
J. Nocedal and S. Wright, \textit{Numerical optimization}. New York, NY,
USA: Springer-Verlag, 2006.

\bibitem{O.El_mmWave_MIMO}
O. El Ayach, S. Rajagopal, S. Abu-Surra, Z. Pi, and R. W. Heath, ``Spatially sparse precoding in millimeter wave MIMO systems,'' \textit{IEEE
Trans. Wireless Commun.}, vol. 13, no. 3, pp. 1499--1513, Mar. 2014.

\end{thebibliography}
\end{document}